\documentclass{article}

\usepackage{arxiv}
\usepackage{comment}
\usepackage[utf8]{inputenc} 
\usepackage[T1]{fontenc}    
\usepackage{hyperref}       
\usepackage{url}            
\usepackage{booktabs}       
\usepackage{amsfonts}       
\usepackage{nicefrac}       
\usepackage{microtype}      
\usepackage{lipsum}
\usepackage{graphicx}
\usepackage{amsmath}
\usepackage{float}
\usepackage{mathtools}
\usepackage{bm}
\usepackage{algorithm}
\usepackage{cleveref}
\usepackage[noend]{algpseudocode}
\usepackage{graphicx}
\usepackage{subcaption}
\usepackage{mwe}
\usepackage{color,soul}
\usepackage{mathtools}

\newcommand*\samethanks[1][\value{footnote}]{\footnotemark[#1]}

\newcommand*{\affaddr}[1]{#1} 

\newcommand*{\email}[1]{\texttt{#1}}
\graphicspath{ {./figures/} }

\title{Integrating Material Selection with Design Optimization via Neural Networks}

\author{%
Aaditya Chandrasekhar \thanks{Contributed equally}, Saketh Sridhara \samethanks,   Krishnan Suresh \thanks{Corresponding author}\\
\affaddr{Department of Mechanical Engineering, University of Wisconsin-Madison}\\
\email{\{achandrasek3, ssridhara, ksuresh\}@wisc.edu%
}}

\begin{document}
\maketitle

\begin{abstract}

The engineering design process  often entails optimizing the underlying geometry while simultaneously selecting a suitable material. For a certain class of simple problems, the two are separable where, for example, one can first  select an optimal material, and then optimize the geometry. However, in general, the two are not separable. Furthermore, the discrete nature of material selection is not compatible with  gradient-based geometry optimization, making simultaneous optimization  challenging.

In this paper, we propose the use of variational autoencoders (VAE) for simultaneous optimization. First, a  data-driven VAE is used to project the discrete material database onto a continuous and differentiable latent space. This is then coupled with a fully-connected  neural network, embedded with a finite-element solver, to simultaneously  optimize the material and geometry. The neural-network's built-in gradient optimizer and back-propagation are exploited during optimization.

The proposed framework is demonstrated using trusses, where an optimal material needs to be chosen from a database, while simultaneously optimizing the cross-sectional areas of the truss members. Several numerical examples illustrate the efficacy of the proposed framework. The Python code used in these experiments is available at \href{https://github.com/UW-ERSL/MaTruss}{github.com/UW-ERSL/MaTruss}

\end{abstract}

\keywords{Material selection \and Truss optimization \and  Ashby charts \and Neural networks \and Autoencoder }
\section{Introduction}
\label{sec:intro}

\begin{figure*}[h!]
	\begin{center}
		\includegraphics[scale= 0.85,trim={0 0 0 0},clip]{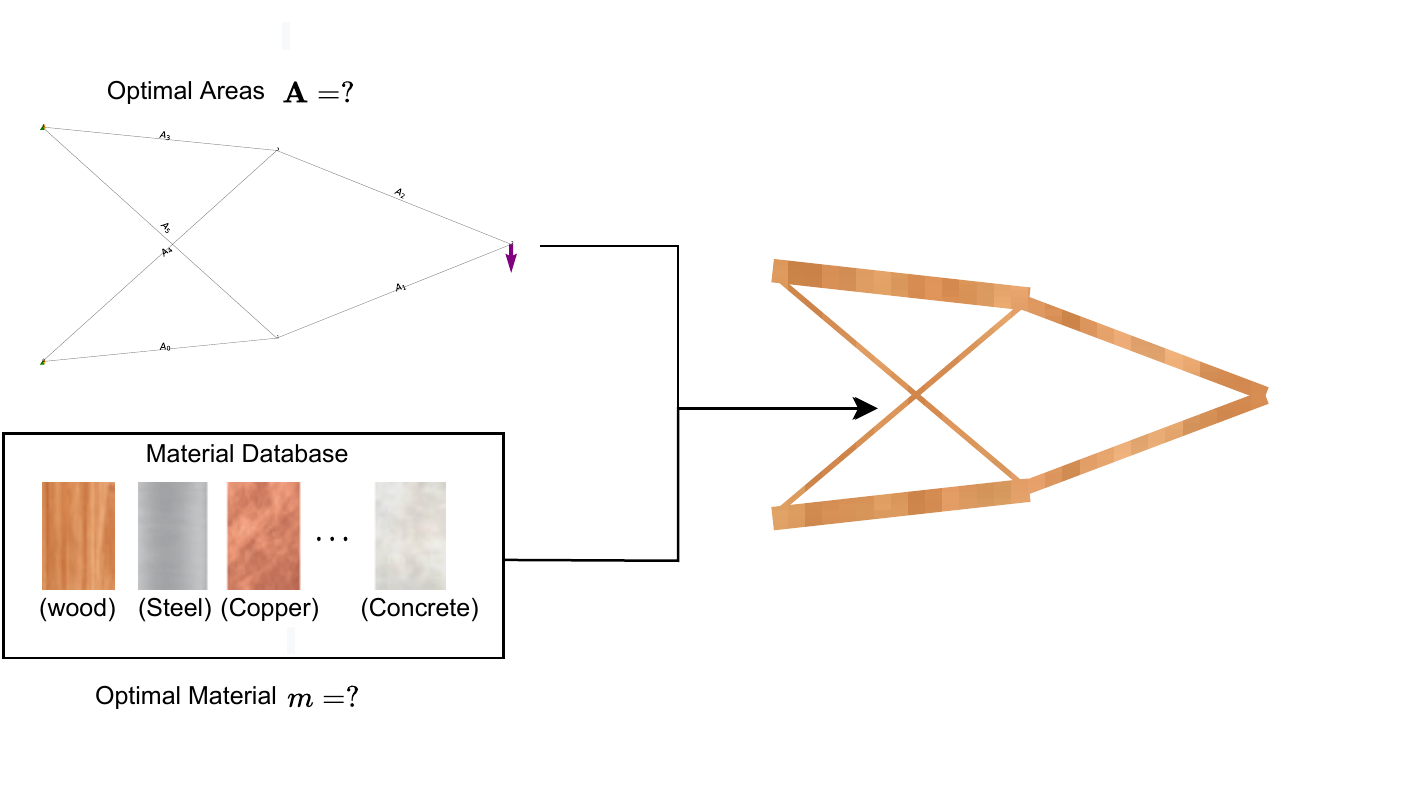}%
		\caption{A truss design problem involving optimizing the cross-sectional areas of truss members, and selecting an optimal material.}
		\label{fig:problem_to_solve}
	\end{center}
\end{figure*}

In engineering design, we are often faced with the task of optimizing the underlying geometry and selecting an optimal material \cite{eggert2005engineering}.  As a simple example, consider the truss design  problem illustrated in \cref{fig:problem_to_solve} where one must optimize the cross-sectional areas, while selecting, for simplicity, a single material for all the truss members. Formally, one can pose this as \cite{rozvany1995layout, achtziger1996truss}:
\begin{subequations}
    \label{eq:optimization_base_Eqn}
	\begin{align}
		& \underset{\bm{A} = \{A_1,A_2,\ldots,A_{N}\},m \in M}{\text{minimize}}
		& & \phi (\bm{A}, \zeta_m) \label{eq:optimization_base_objective}\\
		& \text{subject to}
		& & g (\bm{A}, \zeta_m) \leq 0 \label{eq:optimization_costCons} \\
		& & & K(\bm{A}, E_m)\bm{u} = \bm{f}\label{eq:optimization_base_govnEq}\\
		& & & \bm{A}_{min} \leq \bm{A} \leq  \bm{A}_{max} \label{eq:opt_area_limits}
	\end{align}
\end{subequations}

where $\phi$ is the objective (for example, compliance), $g$ denotes a set of constraints (such as yield stress and buckling), $\bm{f}$ is the applied force, $K$ is the truss stiffness matrix, and $\bm{u}$ is the nodal displacements. The design variables are the cross-sectional areas of the members $\bm{A} = \{A_1, A_2,\ldots,A_{N}\}$, with limits $\bm{A}_{min}$ and $\bm{A}_{max}$, and the material choice $m \in M$.   We denote the collection of relevant material properties by $\zeta_m$; these may include, for example, the Young's modulus ($E_m$), cost per unit mass ($C_m$), mass density ($\rho_m$) and yield strength  ($Y_m$).

\subsection{Literature Review}
 
Observe that the two sets of design variables are tightly coupled. In other words, one cannot, for example, pick a material, and then optimize the cross-sectional areas (or, vice-versa).  
To quote \cite{rakshit2008simultaneous} \textit{"... either approach does not guarantee the optimal combination of geometry and material"}.
Furthermore, while the cross-sectional areas  are  continuously varying, the material choice is discrete, making the problem difficult to solve using  classic gradient-based optimization. \\

If the geometry is fixed, Ashby's method \cite{Ashby1993MaterialSelection} that relies on a \textit{performance metric}, is the most popular strategy for material selection. Furthermore, for simple design problems where the loads, geometry and material functions are separable \cite{ashby2013materials},  a \textit{material index} can be used to select the best material.  This  is simple, efficient and reliable. However, when there are multiple objectives or constraints, a weighted approach is recommended \cite{ashby2000multi}. In practice, computing these weights is not trivial, and the material choice may be far from optimal. Several non-gradient methods have been proposed for material selection \cite{jahan2010material, rao2006material, zhou2009multi}, but these cannot  be integrated with gradient-based  optimization. 
\\

An alternate concept of  a \textit{design index} was proposed in  Ananthasuresh and Ashby \cite{ananthasuresh2003concurrent}. It was shown that the design index can be used to construct a smooth material function, enabling  simultaneous optimization of the geometry and selection of the material via gradient based optimization \cite{rakshit2008simultaneous}.  However, design indices were limited to simple determinate trusses. In \cite{stolpe2004stress}, for the special case of a single stress constraint and a single load, the authors concluded that utmost two materials are sufficient and that the truss design problem can be cast as a linear programming problem.  Unfortunately, this method does not apply when there are multiple constraints.  Recently, the authors of \cite{ching2021truss} considered two materials (glue-laminated timber and steel) to design truss structures incorporating stress constraints as a mixed-integer quadratic problem (MIQPs). However the authors note that   \textit{"[MIQPs]... require algorithmic tuning and considerable computational power"}. Another hybrid approach was proposed in \cite{roy2021hybrid} where a combination of the gradient-based methods such as sequential quadratic programming (SQP)  and an evolution method such as genetic algorithm (GA)  are used to solve MDNLPs. But these methods can be prohibitively expensive.\\

Mathematically, the problem of (discrete) material selection and (continuous) area optimization can be cast as mixed-discrete nonlinear programming problems (MDNLPs). Such problems are fairly common in engineering, and several solution strategies methods have been proposed; see \cite{arora1994methods, martins2021engineering, lee2011mixed}. However, regardless of the strategy, these methods entail repeated solution of a sequence of nonlinear programming problems with careful relaxations and approximations, making them sensitive to assumptions and underlying models \cite{koppe2012complexity}. Furthermore, the popular \emph {branch and bound} algorithm used in solving MDNLPs \cite{martins2021engineering} does not apply here since the optimization problem depends indirectly on the material index through the database.

\subsection{Paper Overview}
\begin{figure*}[h!]
	\begin{center}
		\includegraphics[scale= 0.85,trim={0 0 0 0},clip]{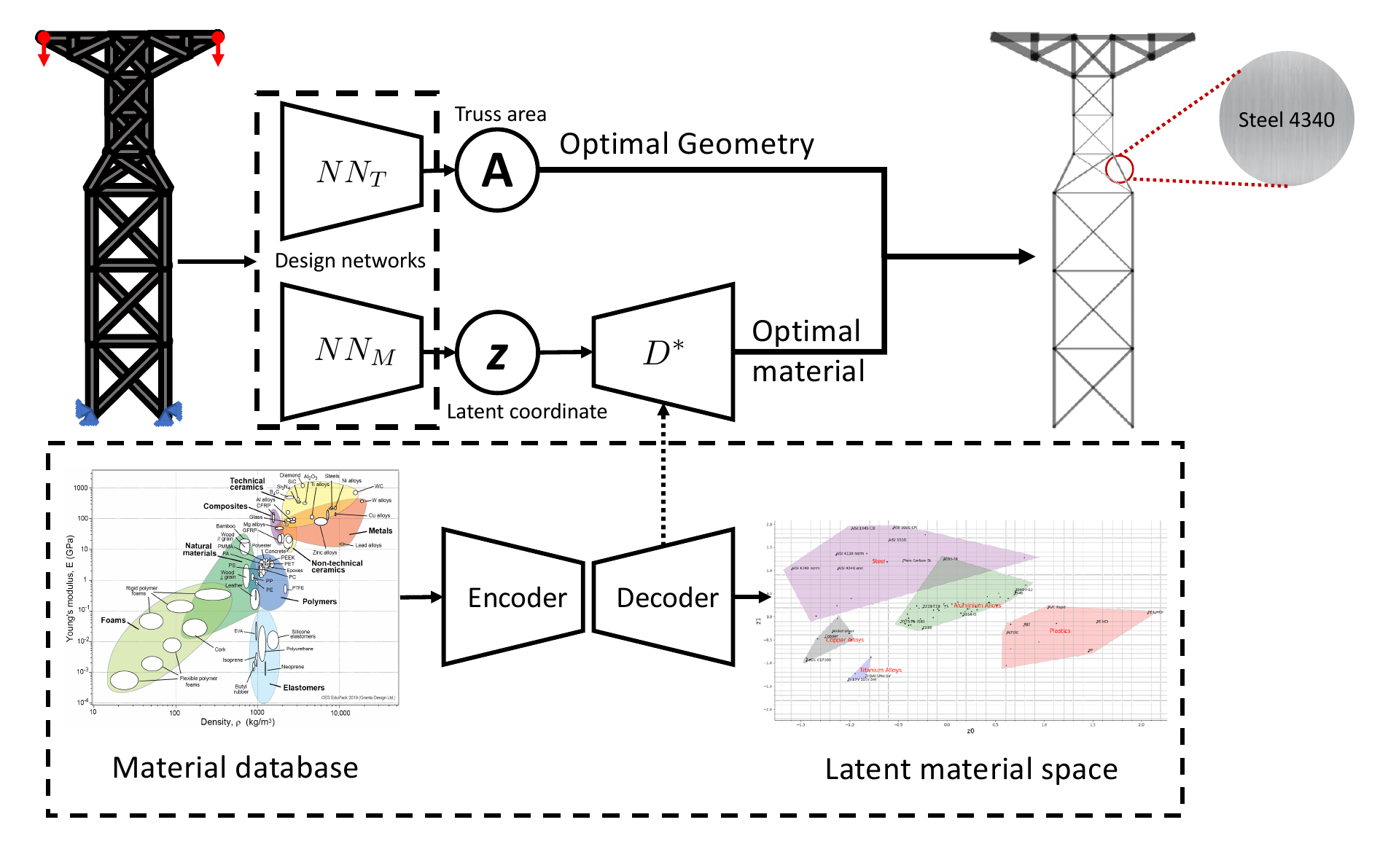}%
		\caption{An overview of the proposed method. }
		\label{fig:graphicalAbstract}
	\end{center}
\end{figure*}

The primary contribution of this work is the use of variational autoencoders (VAEs) to  solve such  problems. VAEs are a special form of neural networks that can convert discrete data, such as a material database, into a continuous and differentiable representation, making the design problem amenable to gradient-based optimization. In the proposed method (see \cref{fig:graphicalAbstract}), a data-driven VAE is trained using a material database, and the trained decoder is then integrated with neural networks to simultaneously optimize the geometry and material.  \\

 The proposed VAE framework is discussed in \cref{sec:differentiableMaterialRepresentation}.  In \cref{sec:designOptimization}, the VAE is integrated with two additional and simple neural networks to solve the truss design problem. Specifically, we leverage the differentiable material representation to simultaneously optimize the geometry and material. Section \ref{sec:expts} demonstrates the proposed framework through several numerical examples. In \cref{sec:conclusion}, limitations and future work are discussed.
\section{A Differentiable Material Representation}
\label{sec:differentiableMaterialRepresentation}
In this section, we discuss how  variational autoencoders (VAEs) can be used to obtain a  continuous and differentiable representation of a discrete material database. This will serve as a foundation for the next section on design optimization.

\subsection{VAE Architecture and Training}
\label{sec:matNet_VAE_cons}
VAEs are popular generative models that have  wide applicability in data compression, semi-supervised learning and data interpolation; please see \cite{kingma2019VAE}. In essence, a VAE captures the data in a form that can be used to synthesize new samples similar to the input data. For example, VAEs have been used to generate new microstructures from image databases \cite{wang2020deep}, in designing photonic crystals \cite{li2020designing},  and heat-conduction materials \cite{guo2018indirect}.

\begin{figure*}[h!]
	\begin{center}
		\includegraphics[scale= 1.,trim={0 0 0 0},clip]{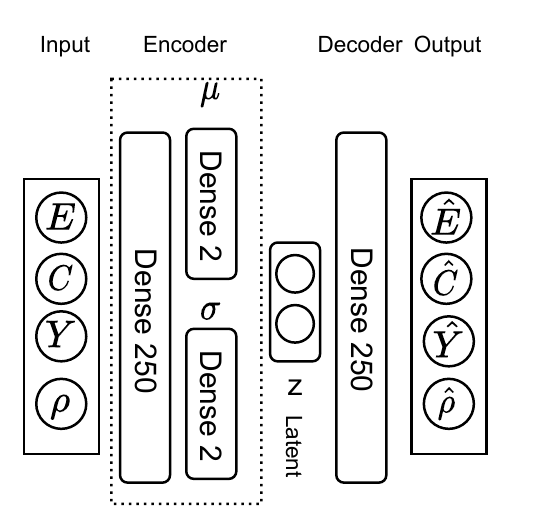}%
		\caption{Architecture of the variational autoencoder.}
		\label{fig:autoencoder_setup}
	\end{center}
\end{figure*}

However, in this paper, we do not use VAEs to synthesize new data, instead \emph {we simply use the VAE's ability to map uncorrelated data onto an abstract latent space}. In that sense, VAEs are similar to principal component analysis (PCA) in its ability  to extract useful information from the data. However, the  nonlinear nature of VAEs allows for far greater generalization than PCA \cite{Goodfellow2016_deepLearning}. In particular, the proposed VAE architecture for capturing material properties is illustrated in \cref{fig:autoencoder_setup}, and consists of the following components:
\begin {enumerate}
\item A four-dimensional \emph {input} module corresponding to the four properties in \cref{table:subsetMaterialList}, namely the Young's modulus ($E$), cost ($C$), mass density ($\rho $) and yield strength ($Y $). The input set is denoted by $\bm{\zeta}$.
\item An \emph {encoder} $F$ consisting of a fully-connected network of 250 neurons, where each neuron is associated with an ReLU activation function and weights \cite{schmidhuber2015deepLearningReview}. 
\item A two-dimensional \emph {latent space}, denoted by ${z_0, z_1}$ that lies at the heart of the VAE.
\item A \emph {decoder} $D$, which is similar to the encoder, consists of a fully-connected network of  250 neurons.
\item A four-dimensional \emph {output} corresponding to the same four properties; the output set is denoted by  $ \hat{\bm{\zeta}}$.
\end{enumerate} 

In this work, the VAE was trained on a material database consisting of 92 materials \cite{SOLIDWORKS:2021}, where \cref{table:subsetMaterialList} represents a small sample.

\label{sec:method_dataDrivenMaterialNet}
\begin{table}[h!]
\centering
\begin{tabular}{|c|c|c|c|c|c|c|c|}
\hline
\textbf{Material} & Class & E $[Pa]$& cost $C[\$/kg]$ & $\rho [kg/m^3]$& $Y [Pa]$ \\ \hline
A286 Iron      & Steel    & 2.01E+11      & 5.18E+00  & 7.92E+03    & 6.20E+08      \\ \hline
AISI 304       & Steel    & 1.90E+11      & 2.40E+00  & 8.00E+03    & 5.17E+08      \\ \hline
Gray Cast Iron & Steel    & 6.62E+10      & 6.48E-01  & 7.20E+03    & 1.52E+08      \\ \hline
3003-H16       & Al Alloy & 6.90E+10      & 2.18E+00  & 2.73E+03    & 1.80E+08      \\ \hline
5052-O         & Al Alloy & 7.00E+10      & 2.23E+00  & 2.68E+03    & 1.95E+08      \\ \hline
7050-T7651     & Al Alloy & 7.20E+10      & 2.33E+00  & 2.83E+03    & 5.50E+08      \\ \hline
Acrylic        & Plastic  & 3.00E+09      & 2.80E+00  & 1.20E+03    & 7.30E+07      \\ \hline
ABS            & Plastic  & 2.00E+09      & 2.91E+00  & 1.02E+03    & 3.00E+07      \\ \hline
PE HD          & Plastic  & 1.07E+09      & 2.21E+00  & 9.52E+02    & 2.21E+07      \\ \hline
\end{tabular}
\caption{A curated subset of materials and their properties used in the training.}
\label{table:subsetMaterialList}
\end{table}

\begin{figure*}[h!]
	\begin{center}
		\includegraphics[scale= 0.75,trim={0 0 0 0},clip]{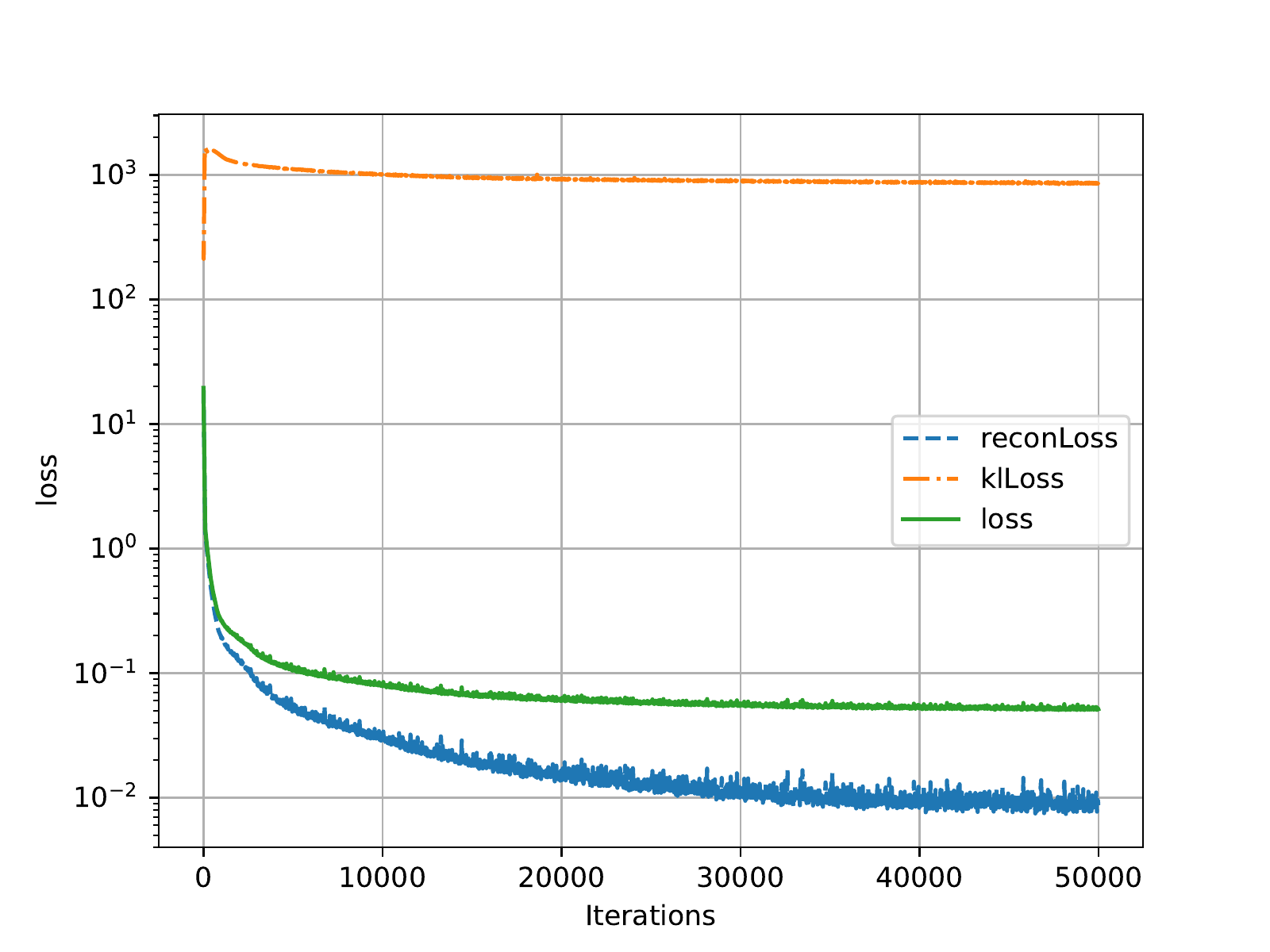}%
		\caption{Convergence plot during training of the VAE.}
		\label{fig:convergenceVAE}
	\end{center}
\end{figure*}

As mentioned earlier, the VAE's primary task is to match the output to the input as closely as possible. This is done through an optimization process (also referred to a training), using the weights associated with the encoder and decoder as optimization parameters. In other words, we minimize $ ||\bm{\zeta} - \hat{\bm{\zeta}}|| $. Additionally, a KL divergence loss is imposed to ensure that latent space  resembles a standard Gaussian distribution $z \sim \mathcal{N}(\mu = 0 , \sigma = 1)$ \cite{kingma2019VAE}. Thus, the net loss can then  expressed as:
\begin{equation}
    L = ||\bm{\zeta} - \hat{\bm{\zeta}}|| + \beta \;  KL(z || \mathcal{N})
    \label{eq:loss_VAE}
\end{equation}
where $\beta$ is set to a recommended value of $5*10^{-5}$. Further the input is scaled  between $(0,1)$ for all attributes, and the output is re-scaled back after training. This ensures that all material properties are weighted equally. In this work, we used PyTorch \cite{NEURIPS2019_9015_pyTorch} to model the VAE, and the gradient-based Adam optimizer \cite{Kingma2015ADAM} to minimize \cref{eq:loss_VAE} with a learning rate of $0.002$ and $50,000$ epochs. The convergence  is illustrated in \cref{fig:convergenceVAE}; the training took approximately $51$ seconds on a Macbook M1 pro.

\begin{figure*}[h!]
	\begin{center}
		\includegraphics[scale= 0.4,trim={0 0 0 0},clip]{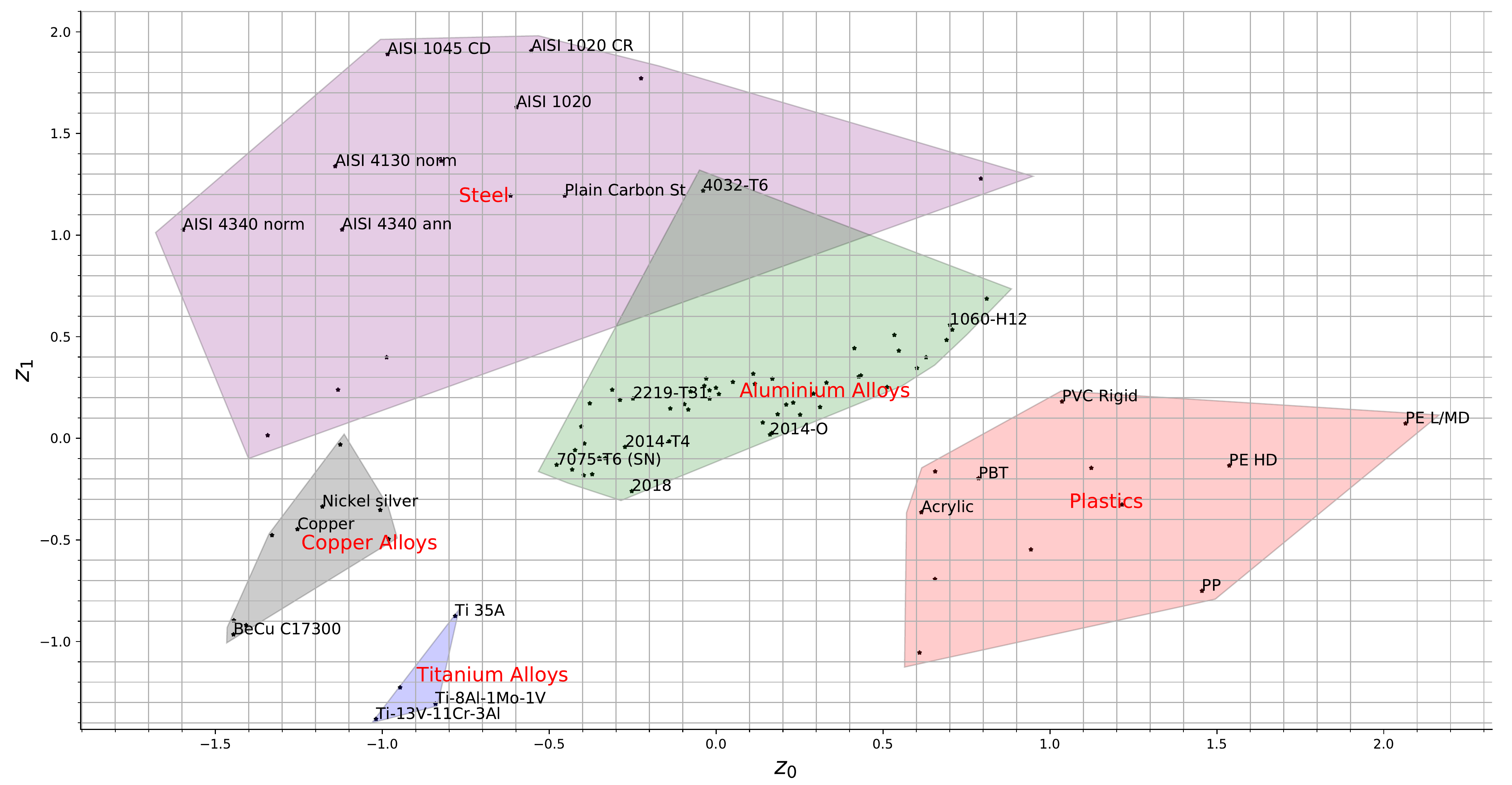}%
		\caption{Encoding of the materials in the latent space; only a few materials are annotated for clarity.}
		\label{fig:latentField}
	\end{center}
\end{figure*}

Thus, the VAE captures each material  uniquely and unambiguously using a non-dimensional latent space, i.e., ${z_0, z_1}$ in our case; this is visualized in \cref{fig:latentField}. For example, annealed AISI 4340 is represented by the pair $(-1.2,1.0)$ while Acrylic is represented by  $(0.6,-0.4)$.  The VAE also clusters similar materials together in the latent space, as can be observed in \cref{fig:latentField}. In an abstract sense, the latent space in \cref{fig:latentField} is similar to the popular Ashby charts \cite{Ashby1993MaterialSelection}.

\subsection{ Representational  Accuracy}
\label{sec:latentspace}
One  can expect the values reconstructed using the decoder to deviate from the true material data; \cref{table:errorMaterialList} summarizes the errors. The following observations are worth noting: 

\begin {enumerate}
\item Despite the lack of correlation between  material properties, and two orders of magnitude difference in values, the VAE captures the entire database of 92 materials reasonably well, using a simple two-dimensional latent space.
\item The error can be further reduced by either increasing the dimension of the latent space, or limiting the type of materials considered (see  \cref{sec:expts}).
\item Finally, as will be discussed in the next section,  the  original material data will be used, as part of a post-processing step, at the end of optimization.
\end {enumerate}

\begin{table}[h!]
\centering
\begin{tabular}{|c|c|c|c|c|}
\hline
\textbf{Material} & \textbf{$\Delta E \%$} & $\Delta C \%$ & \textbf{$\Delta \rho \%$} & $\Delta Y \%$ \\ \hline
A286 Iron    & 0.8 & 0.7 & 0.8 & 0.0 \\ \hline
ABS          & 1.4 & 0.0 & 0.7 & 1.7 \\ \hline
AISI 304     & 4.3 & 3.1 & 0.4 & 1.0 \\ \hline
Gray Cast Fe & 1.3 & 1.8 & 0.2 & 0.2 \\ \hline
3003-H16     & 3.2 & 6.8 & 1.3 & 1.1 \\ \hline
5052-O       & 1.4 & 2.4 & 0.6 & 1.2 \\ \hline
7050-T7651   & 0.9 & 1.9 & 1.0 & 1.9 \\ \hline
Acrylic      & 2.3 & 0.5 & 0.2 & 0.5 \\ \hline
PE HD        & 0.7 & 0.9 & 1.1 & 1.8 \\ \hline
\textbf{Max error}     & \textbf{5.0}           & \textbf{6.8}  & \textbf{3.3}              & \textbf{7.1}  \\ \hline
\end{tabular}
\caption{Percentage error between actual and decoded data.}
\label{table:errorMaterialList}
\end{table}
 
\subsection{Differentiability of Latent Space}
\label{sec:vaeNet_differentiability}

A crucial aspect of the latent space is its differentiability. In particular, we note that using the decoder, each of the output material properties  is represented via analytic activation functions; for example, $ \hat{E}  = D^*_E(z_0, z_1)$ where $D^*_E(\cdot)$ is a combination of activation functions and weights. It means that one can  back propagate through the decoder to compute analytical sensitivities; for example, the sensitivity ($\frac{\partial \hat{E}}{\partial z_0}$). This plays an important role in  gradient-driven design optimization.

\subsection{Material Similarity}
\label{sec:matNet_interpretLatentSpace}

\begin{figure*} []
	\begin{center}
		\includegraphics[scale= 0.4,trim={0 0 0 0},clip]{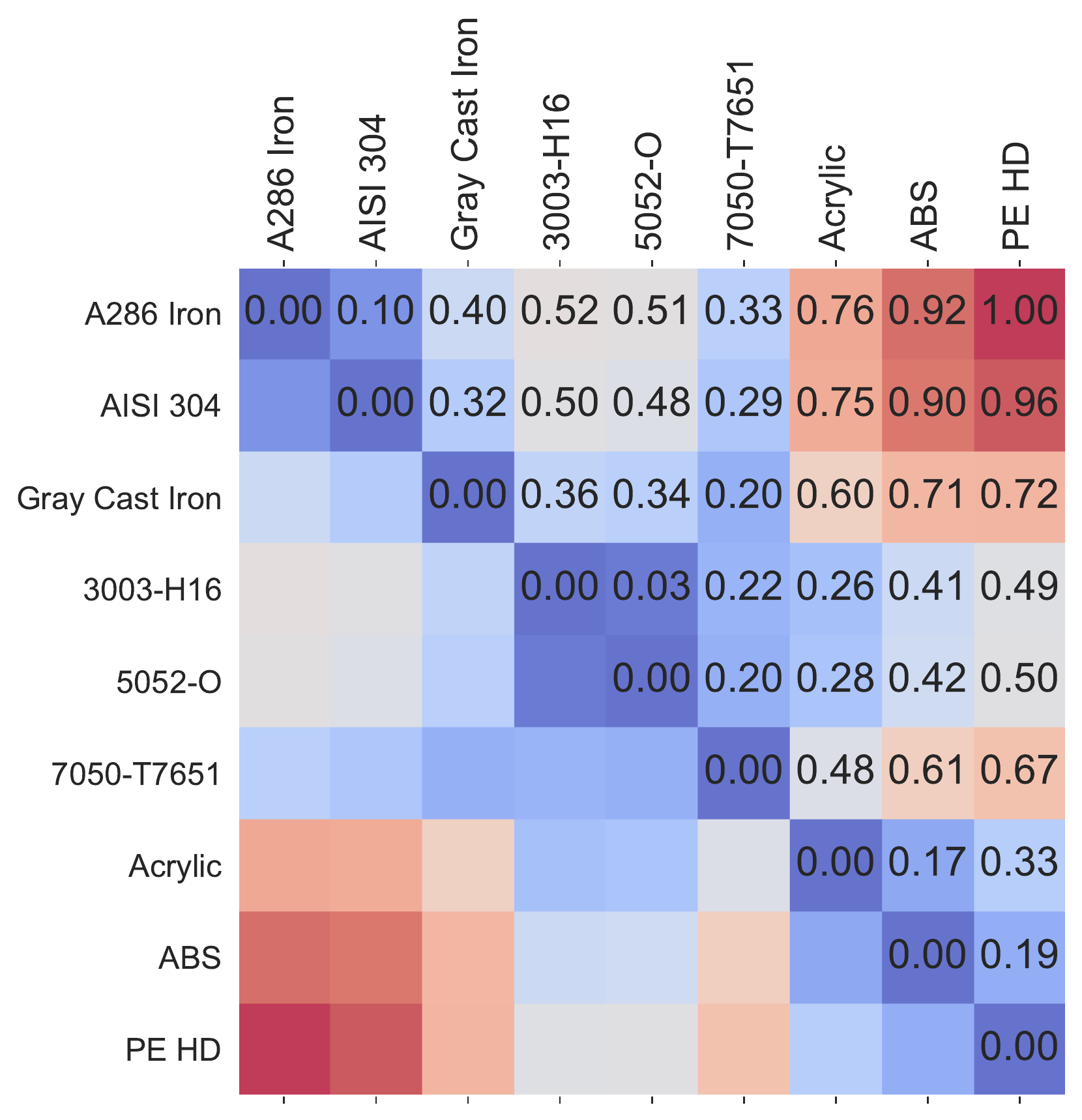}%
		\caption{Symmetric distance in the latent space between materials}
		\label{fig:distanceMatrix}
	\end{center}
\end{figure*}

While the accuracy and differentiability of the VAE lends itself to design optimization, the  latent space also provides key insights into material characteristics. For instance, one can compute the Euclidean distance between materials in the latent space. These distances are reported in  \cref{fig:distanceMatrix}. As one can expect, steels are closer to aluminum alloys, than they are to plastics. 

\begin{figure*} []
        \centering
        \includegraphics[scale=0.425]{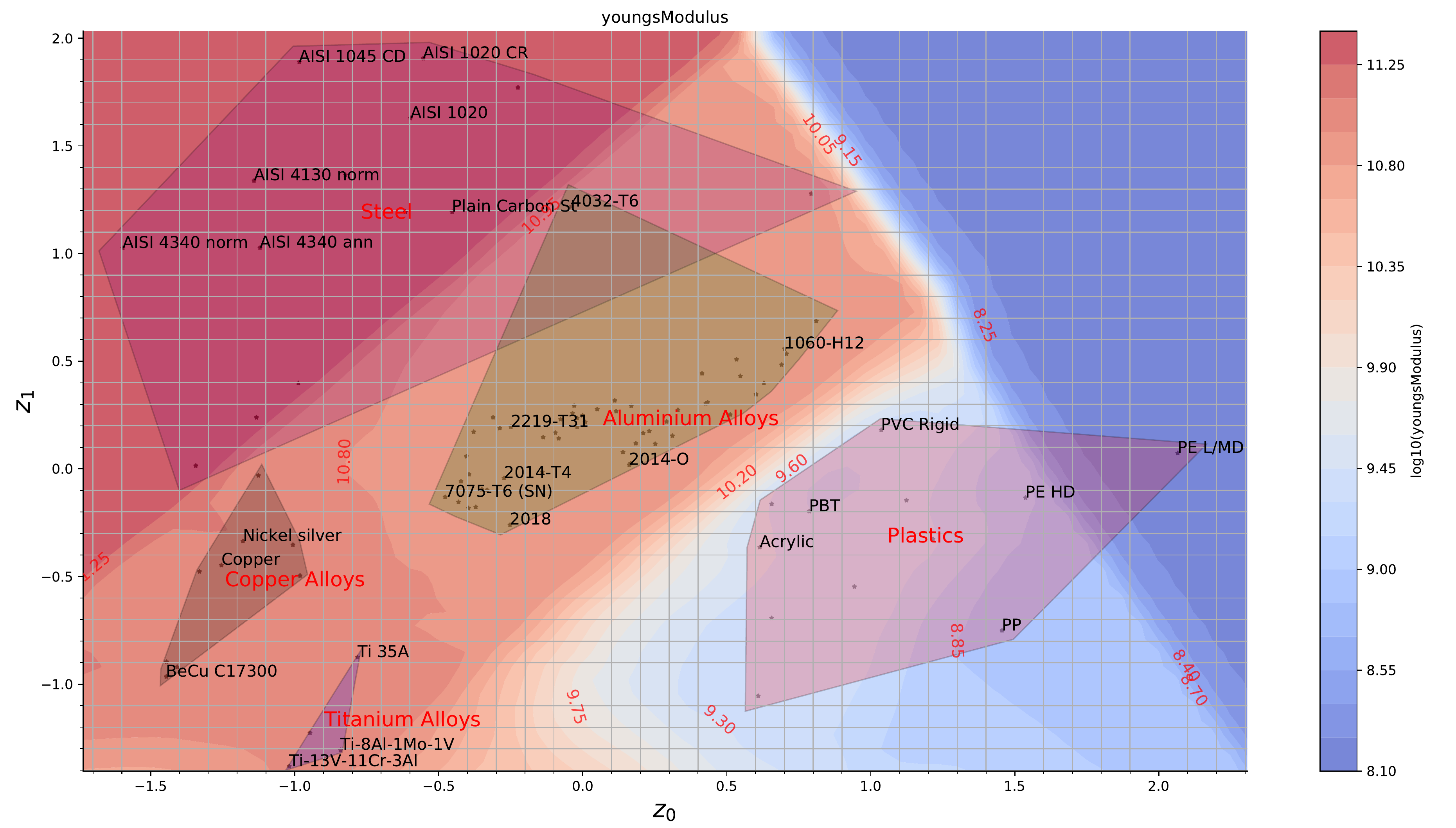}
        \caption{Contours of the Young's moduli in the latent space. } 
        \label{fig:latentFieldYoungsModulus}
    \end{figure*}

Furthermore, one can overlay the latent space in  \cref{fig:latentField} with specific material properties to gain further insight. This is illustrated in  \cref{fig:latentFieldYoungsModulus} where the contour plots of the Young's moduli are illustrated in the latent space. This allows designers to visualize similarities between materials, specific to a material property.

\section{Design Optimization}
\label{sec:designOptimization}
Having constructed a differentiable representation of material properties, one can now pose the optimization problem discussed in \cref{sec:intro}, now using  $\{\bm{A}, z_0, z_1\}$ as continuous design variables. Without a loss in generalization,  we consider a specific instance of the truss optimization problem, where the objective is the compliance with three sets of constraints: cost constraint, buckling constraint, and tensile yield constraint. We assume here that members will fail due to buckling, before failing due to compressive yield \cite{rakshit2008simultaneous}. The resulting optimization problem for circular truss members can be posed as \cite {suresh_2021}:
\begin{subequations}
	\label{eq:optimization_nn_Eqn}
	\begin{align}
		& \underset{\bm{A} = \{A_1,A_2,...A_{N}\},z_0, z_1}{\text{minimize}}
		& &J = \bm{f}^T \bm{u}(\bm{A},\hat{E}) \label{eq:optimization_nn_objective}\\
		& \text{subject to}
		& & [K(\bm{A},\hat{E})]\bm{u} = \bm{f}\label{eq:optimization_nn_govnEq}\\
		& & & g_c \coloneqq
		\bigg(\frac{\hat{\rho} \hat{C}}{C^*} \sum\limits_{k=1}^{N} A_k L_k\bigg) - 1 \leq 0 \label{eq:optimization_nn_costCons} \\
		& & & g_b \coloneqq \underset{k}{\text{max}} \big( \frac{-4 P_{k} L_k^2}{\pi^2 \hat{E} A_k^2} \big) - \frac{1}{F_s}  \leq  0 \label{eq:optimization_buckling_nn_cons} \\
        & & & g_y \coloneqq \underset{k}{\text{max}} \big( \frac{P_{k}}{\hat{Y} A_k} \big) - \frac{1}{F_s} \leq 0 \label{eq:optimization_nn_yieldingCons} \\
		& & & \bm{A}_{min} \leq \bm{A} \leq \bm{A}_{max} \label{eq:opt_nn_area_limits}
	\end{align}
\end{subequations}
where  $P_k$ is the internal force in member $k$, $L_k$ is its length, $C^*$ is the allowable cost, $F_s$ is the safety factor, and the material properties ($\hat{E}, \hat{\rho}, \hat{Y}, \hat{C}$) are decoded from the latent space coordinates ($z_0,z_1$).  Further, to facilitate gradient driven optimization, the $max$ operator is relaxed here using a p-norm, i.e., $\underset{i}{\text{max}}(x_i) \approx || \bm{x}||_p$, with $p=6$ . 

\begin{figure}[]
     \centering
     \begin{subfigure}[t]{0.3\textwidth}
         \centering
         \includegraphics[scale = 1]{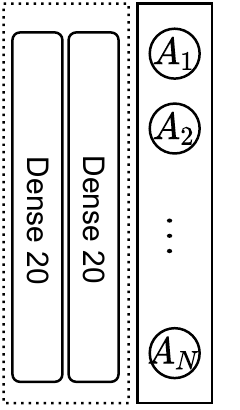}
         \caption{Topology network $NN_T$}
         \label{fig:topologyNetwork}
     \end{subfigure}
     \hfill
     \begin{subfigure}[t]{0.4\textwidth}
         \centering
         \includegraphics[scale=0.75]{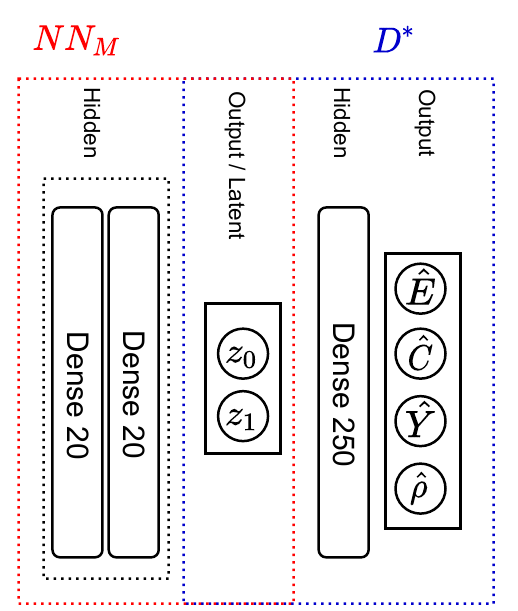}
         \caption{Material network $NN_M$ with  decoder.}
         \label{fig:materialNetworkInterfacedWithDecoder}
     \end{subfigure}
    \caption{The truss and material neural networks.}
    \label{fig:designNetworks}
\end{figure}

To solve the above optimization we use two additional neural networks (NNs) (see \cref{fig:topologyNetwork}): a truss network $NN_T$, parameterized by weights $\bm{w}_T$, and a material network $NN_M$, parameterized by weights $\bm{w}_M$. The truss network $NN_T$  is a simple feed-forward NN with two hidden layers with a width of 20 neurons, each containing an ReLU activation function. The output layer of $NN_T$ consists of $N$ neurons where $N$ is the number of truss members, activated by a Sigmoid function, generates a vector $\bm{O_T}$  of size $N$ whose values are in $[0,1]$. The output is then scaled as $\bm{A} \leftarrow \bm{A}_{min} + \bm{O_T}(\bm{A}_{max} - \bm{A}_{min})$ to satisfy the area bounds \cref{eq:opt_nn_area_limits}. The material network  $NN_M$  (\cref{fig:materialNetworkInterfacedWithDecoder}) is similar to  to $NN_T$, except that the output layer consists of  two output neurons activated by Sigmoid functions. The outputs  $\bm{O_M}$ are scaled as $\bm{z} \leftarrow -3 + 6 \bm{O_M}$, resulting in $z_i \in [-3,3]$ corresponding to six Gaussian deviations. These outputs now interface with the trained decoder $D^*$ from \cref{fig:autoencoder_setup} . Thus, by varying the weights $\bm{w}_M$ of the material network one can create points in the latent space, that then feeds to the trained decoder resulting in values of material constants.

\subsection{Loss Function}
\label{sec:designOptimizationLoss}

\begin{figure*} [h!]
        \centering
        \includegraphics[scale=0.85,trim={140 0 0 0},clip]{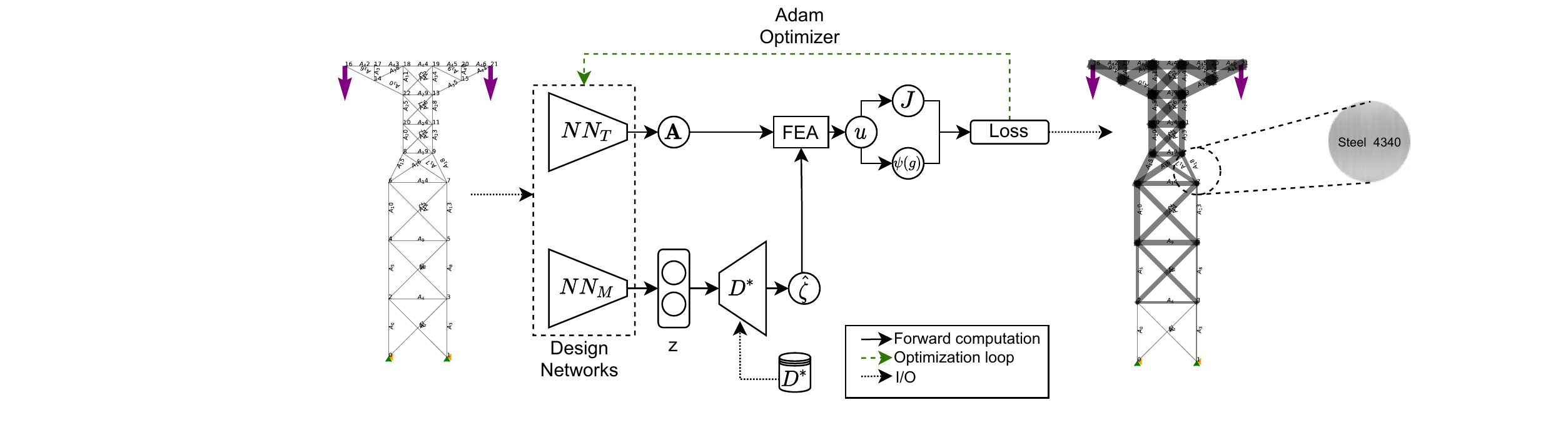}
        \caption{Forward computation} 
        \label{fig:fwdMaTrussFlowchart}
    \end{figure*}

With the introduction of the two  NNs,  the weights $\bm{w}_T$  now control the areas $\bm{A}$, while the weights  $\bm{w}_M$ control the material constants $\hat{\bm{\zeta}}$. In other words, the weights $\bm{w}_T$  and $\bm{w}_M$ now form the design variables. Further, since NNs are designed to minimize an unconstrained loss function, we  convert the constrained minimization problem in \cref{eq:optimization_nn_Eqn} into an unconstrained minimization by employing a log-barrier scheme as proposed in  \cite{kervadec2019constrlogBarrierOptML}:
\begin{equation}
    L(\bm{w}_T, \bm{w}_M) = J + \psi(g_c) + \psi(g_b) + \psi(g_y)
    \label{eq:lossFunction}
\end{equation}

where, 
\begin{equation}
    \psi_t(g) = \begin{cases}
    -\frac{1}{t} \log(-g) \; ,\quad g \leq \frac{-1}{t^2}\\
    tg - \frac{1}{t} \log(\frac{1}{t^2}) + \frac{1}{t} \; , \quad \text{otherwise}
    \end{cases}
\end{equation}

with the parameter $t$ updated during each iteration (as described in the next section). Thus the optimization problem reduces to a simple form:

\begin{subequations}
	\label{eq:optimization_adam_Eqn}
	\begin{align}
		& \underset{\bm{w}_T, \bm{w}_M}{\text{minimize}}
		& &L(\bm{w}_T, \bm{w}_M) \label{eq:optimization_adam_objective}\\
		& \text{subject to}
		& & [K(\bm{A}(\bm{w}_T),\hat{E}(\bm{w}_M))]\bm{u} = \bm{f}\label{eq:optimization_adam_govnEq}
	\end{align}
\end{subequations}
A schematic of the proposed framework is presented in \cref{fig:fwdMaTrussFlowchart}.

\subsection{Structural Analysis}
\label{sec:FEA}

We rely on classical structural analysis to evaluate the performance of the truss structure during each iteration.  The solver compute the stiffness matrix for each element and upon assembling the global stiffness matrix, the nodal displacement vector u are computed using a sparse solver  in PyTorch \cite{NEURIPS2019_9015_pyTorch}. This allows us to exploit  backward propagation for automatic differentiation \cite{ChandrasekharAuTO2021}, resulting in an end-to-end differentiable solver with automated sensitivity analysis as described next. 

\subsection{Sensitivity Analysis}
\label{sec:method_physicsDrivenTO_sensAnalysis}

\begin{figure*}[h!]
	\begin{center}
		\includegraphics[scale= 0.85,trim={140 0 0 0},clip]{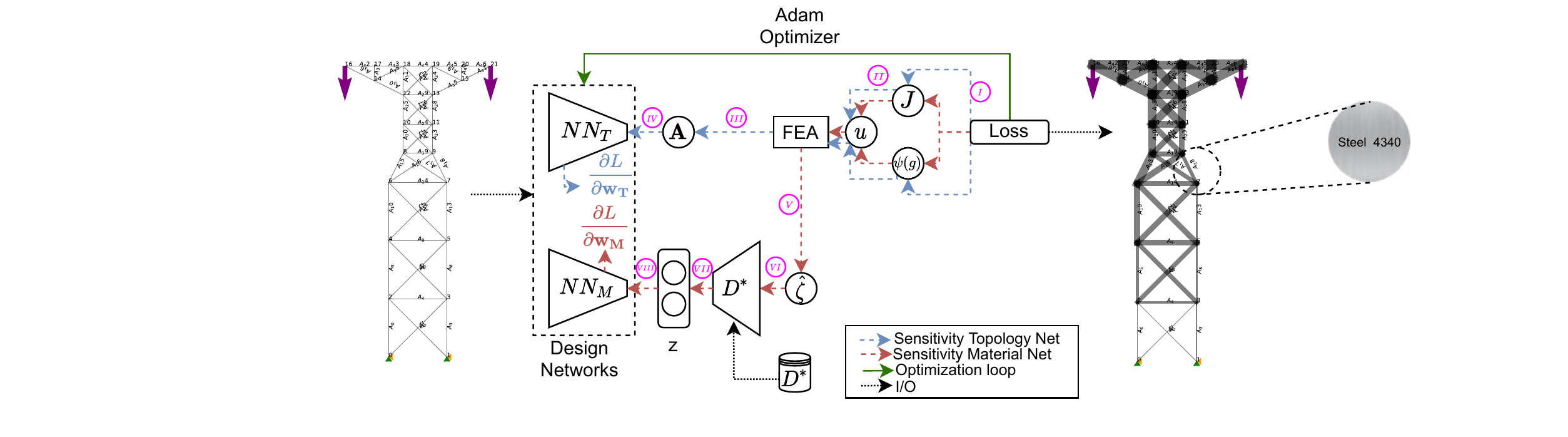}%
		\caption{Backward computation}
		\label{fig:bwdMaTrussFlowchart}
	\end{center}
\end{figure*}

In the proposed framework, the loss function in \cref{eq:lossFunction} is minimized using gradient-based Adagrad optimizer \cite{duchi2011adaptive}. Further, the sensitivities are computed automatically using back propagation. A schematic representation of the backward computation graph is shown in \cref{fig:bwdMaTrussFlowchart}. For instance the sensitivity of the loss function with respect to  the weights $\bm{w}_T$ can be expressed as:
\begin{equation}
    \frac{\partial L}{\partial \bm{w}_T} = \bigg[ \underbrace{\frac{\partial L}{\partial J}}_I\underbrace{\frac{\partial J}{\partial \bm{u}}}_{II} \underbrace{\frac{\partial \bm{u}}{\partial \bm{A}}}_{III} + \ldots \bigg] \underbrace{\frac{\partial \bm{A}}{\partial \bm{w}_T}}_{IV}
    \label{eq:sens_topWts}
\end{equation}

This is illustrated in  \cref{fig:bwdMaTrussFlowchart} where each term corresponds to an edge in the graph. Similarly, the  sensitivity with respect to $\bm{w}_M$ is given by:
\begin{equation}
    \frac{\partial L}{\partial \bm{w}_M} =  \Bigg[ \underbrace{\frac{\partial L}{\partial J}}_{I} \underbrace{\frac{\partial J}{\partial \bm{u}}}_{II} \underbrace{\frac{\partial \bm{u}}{\partial \hat{\zeta}}}_{V} + \ldots \bigg] \underbrace{\frac{\partial \hat{\zeta}}{\partial \bm{w}_{D^*}}}_{VI} \underbrace{\frac{\partial \bm{w}_{D^*}}{\partial z}}_{VII}\underbrace{\frac{\partial \bm{z}}{\partial \bm{w}_M}}_{VIII}
    \label{eq:sens_matWts}
\end{equation}
These are once again illustrated in  \cref{fig:bwdMaTrussFlowchart}. Note that the backward computation is carried out automatically in the proposed framework.

\subsection{Post Processing}
\label{sec:postProcessing}

Once the optimization process is complete, we obtain an optimal set of latent coordinates $\bm {z^*}$ . However, there might not exist a material in the  database corresponding precisely to  $\bm {z^*}$.  We therefore define a \emph {confidence metric} $\gamma_{m}$  for each material $m$ by using the distance from $\bm {z^*}$ to the material $m$ in the latent space:

\begin{equation}
    \gamma(\bm{z}_m, \bm{z}^*) = 1 - \frac{|| \bm{z}^* - \bm{z}_m||}{\underset{\forall k \in M}{\text{max}} (||\bm{z}^* - \bm{z}_k||)}
    \label{eq:confidence}
\end{equation}

The metric serves to rank the materials based on their distance from $\bm {z^*}$.  After finding the closest material, we repeat the geometry optimization to compute the optimal areas $\bm {A^*}$. This is analogous to the \emph {rounding} method in discrete optimization \cite{martins2021engineering}: "\emph {After rounding, it is usually best to repeat the optimization once more, allowing only the continuous design variables to
vary. This process may not lead to the exact optimum, and sometimes may not even lead to a feasible solution, but for many problems this is an effective approach.}" This is explored further through numerical experiments in \cref{sec:expts}.
\subsection{Algorithm}
\label{sec:algo}
The proposed algorithms are summarized in this section. As mentioned earlier, we assume that a  material database has been provided. The procedure begins with training the VAE as described in \cref{alg:encodeMat}.  The encoder is then discarded, and the decoder retained. In  \cref{alg:MaTruss}, starting with the given truss, specified loads, restraints, area bounds and other constraints, the areas and material are optimized. In our experiments $t_0 = 3$ and $\mu = 1.01$, and the learning rate for the Adagrad optimizer was set to the value of $2*10^{-3}$.  Once optimized, we find the optimal areas using the nearest material in the latent space.

\begin{algorithm}[h!]
	\caption{Encode Materials}
	\label{alg:encodeMat}
	\begin{algorithmic}[1]
		\Procedure{MatEncode}{$\zeta$ , $F$, $D$}
		\Comment{Input: Training data, encoder and decoder}

		\State  epoch = 0 \Comment{iteration counter}

		\Repeat \Comment{VAE training}
		
		\State $F(\zeta) \rightarrow \{\mu, \sigma\}$ \Comment{Forward prop. encoder}
		
		\State $\{ \mu,\sigma \} \rightarrow z $ \Comment{Reparameterization  \cite{kingma2019VAE}}
		
		\State $\{ \mu,\sigma \} \rightarrow KL(z||\mathcal{N})$ \Comment{KL loss}
		
		\State $D(z) \rightarrow \hat{\zeta}$ \Comment{Forward prop. decoder}
		
		\State $\{ \zeta, \hat{\zeta}, KL\} \rightarrow L$ \Comment{VAE Loss}
		
		\State $\bm{w} + \Delta\bm{w}(\nabla L) \rightarrow \bm{w} $ \Comment{Update VAE weights, Adam Optimizer}

		\State $\text{epoch}++$
		
		\Until{ error is acceptable} \Comment{Iterate}
		
	    \State	\Return $D$ \Comment{Trained decoder}
		\EndProcedure
	\end{algorithmic}
\end{algorithm}

\begin{algorithm}[h!]
	\caption{Truss Optimization}
	\label{alg:MaTruss}
	\begin{algorithmic}[1]
		\Procedure{TrussOpt}{truss, loads, restraints, $D^*$, ...}
		\Comment{Input: Initial truss, trained decoder, max cost}

		\State  k = 0 \Comment{iteration counter}
		
		\Repeat \Comment{Optimization (Training)}
		
		\State $NN_T(\bm{w}_T) \rightarrow \bm{A} $ \Comment{Fwd prop $NN_T$; compute truss areas}\label{alg:fwdPropNN_T}

		\State $NN_M(\bm{w}_M) \rightarrow z$ \Comment{Fwd prop $NN_M$; compute latent coordinate}\label{alg:fwdPropNN_T}
		
		\State $D^*(z) \rightarrow \hat{\bm{\zeta}}$ \Comment{Fwd prop $D^*$; compute material properties}\label{alg:fwdPropTrainedDcoder}

		\State $\{\bm{A}, \hat{E}\} \rightarrow [K]$ \Comment { compute stiffness matrix}\label{alg:densityToStiffness}
		
		\State $\{[K],\bm{f} \} \rightarrow \bm{u}$  \Comment{State Eq., FEA \cref{eq:optimization_nn_govnEq}}\label{alg:feSolve}
		
		\State $\{\bm{u}, \bm{f} \} \rightarrow J$  \Comment{Compliance, design objective \cref{eq:optimization_nn_objective}}\label{alg:complianceObjective}

		\State $\{ \bm{A}, \hat{\rho}, \hat{C} \} \rightarrow g_c$  \Comment{cost constraint \cref{eq:optimization_nn_costCons}}\label{alg:costCons}
		
		\State $\{ \bm{A}, \hat{E} \} \rightarrow g_b$  \Comment{Buckling constraint \cref{eq:optimization_buckling_nn_cons}}\label{alg:costCons}
		\State $\{ \bm{A}, \hat{Y} \} \rightarrow g_y$  \Comment{Yielding constraint \cref{eq:optimization_nn_yieldingCons}}\label{alg:costCons}
		\State $\{J, g_c, g_b, g_y\} \rightarrow L$ \Comment{Loss \cref{eq:lossFunction}} \label{alg:lossFunction}

		\State $AD(L \leftarrow \bm{w}_T \; , \; g_c \leftarrow \bm{w}_T \; , \; g_b \leftarrow \bm{w}_T, \; g_y \leftarrow \bm{w}_T) \rightarrow \nabla_{\bm{w}_T} L$ \Comment{auto diff. for sens. w.r.t. $NN_T$ \cref{eq:sens_topWts}}\label{alg:sensAnalNN_T}
		
		\State $AD(L \leftarrow \bm{w}_M \; , \; g_c \leftarrow \bm{w}_M, \; g_b \leftarrow \bm{w}_M, \; g_y \leftarrow \bm{w}_M) \rightarrow \nabla_{\bm{w}_M} L$ \Comment{auto diff. for sens. w.r.t. $NN_M$ \cref{eq:sens_matWts}}\label{alg:sensAnalNN_M}
		
		\State $\bm{w}_j + \Delta \bm{w}_j(\nabla_{\bm{w}_j}L) \rightarrow \bm{w}_j \; ; \quad j = \{T,M\}$ \Comment{AdaGrad optimizer step} \label{alg:adamStep}

		\State $ t  \leftarrow t_0\mu^{k}$ \Comment {Increment $t$; log-barrier term} \label{alg:OptPenaltyUpdate}

		\State $\text{k}++$
		
		\Until{ $|| \Delta w || < \epsilon^*$ } \Comment{Check for convergence}
		
	    \State	\Return $\{ z, \bm{A} \}$
	    
	    \State Find nearest material and optimize for area \Comment{\cref{sec:postProcessing}}
		\EndProcedure
	\end{algorithmic}
\end{algorithm}

\section{Numerical Experiments}
\label{sec:expts}

In this section, we present results from several experiments that illustrate the method and algorithm. 

\subsection{Simultaneous Optimization of Area and Material}

\begin{figure}[h!]
    \centering
    \includegraphics[scale = 0.35]{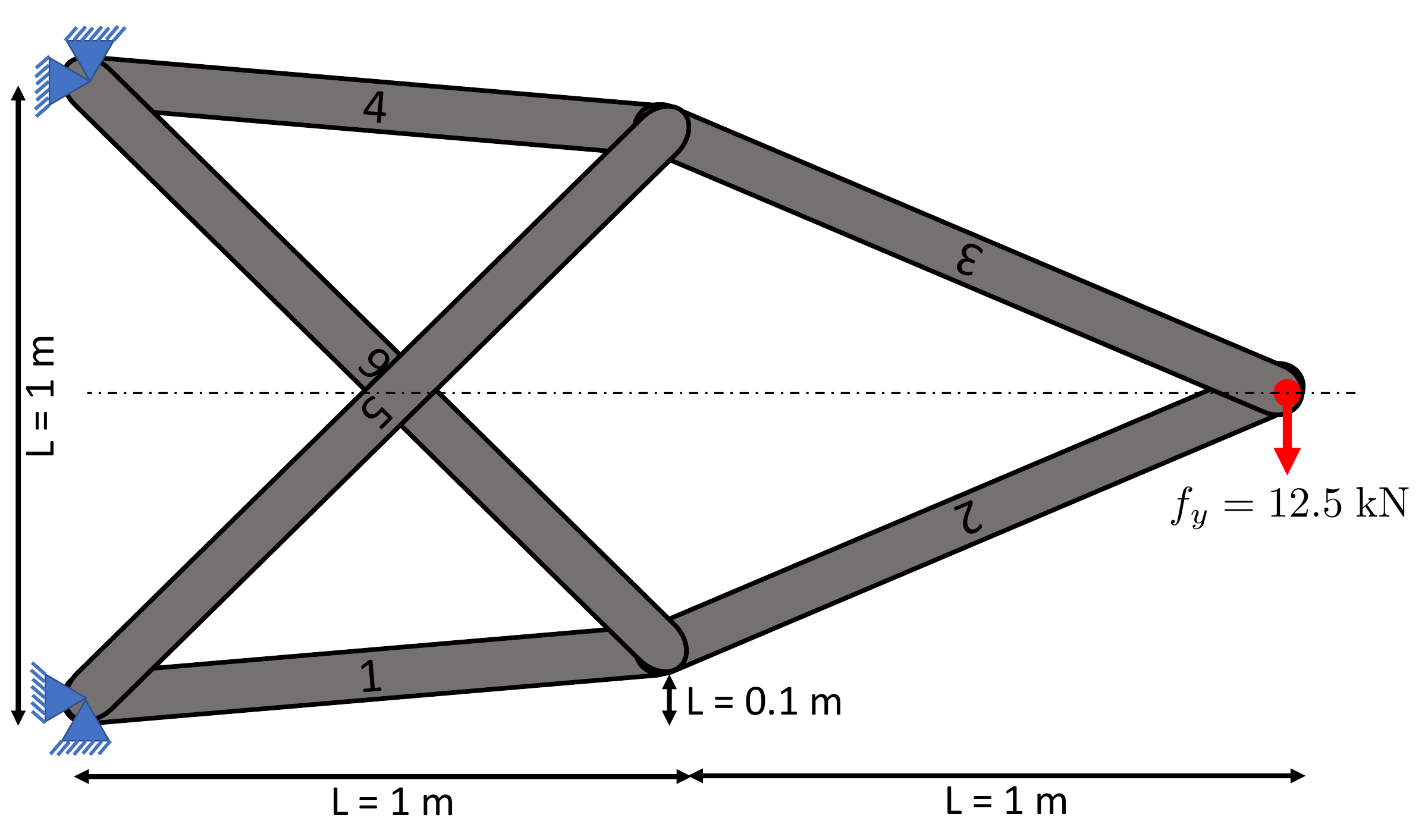}
    \caption{Loading of 6-bar mid-cantilever truss.}
    \label{fig:midCantBC}
\end{figure}

The central hypothesis of this paper is that simultaneous optimization of material and geometry leads to superior performance compared to sequential optimization. To validate this hypothesis, we consider the  6-bar truss system illustrated in \cref{fig:midCantBC}, and perform the following three experiments:

\begin{enumerate}
    \item \textbf{Material optimization}: Here the areas are fixed at the initial value of $A_k = 2*10^{-3}  m^2,  \forall k$, and only the material is optimized using the decoder. 
    \item \textbf{Area optimization}: Using the optimal material from the first experiment,  the areas are optimized by disregarding the decoder.
    \item \textbf{Simultaneous material and area optimization}: Here the material and areas are optimized simultaneously.
\end{enumerate}

The results are reported in \cref{table:perf_threeCases_costCons_midCant} for a cost constraint of $C^* = \$60$ and a safety factor $F_S = 4$. Observe that scenario-1 (material optimization) leads to the lowest performance since the areas are not optimized. For this scenario, we also performed a brute-force search wherein the compliance was evaluated for all the materials; the brute-force search also resulted in `AISI 4130 norm' as the best choice. In scenario-2, when the areas are optimized using `AISI 4130 norm' as the material, the performance improved significantly.  Finally, the best performance is achieved in scenario-3 when both material and areas are optimized simultaneously. Observe that the optimal material in scenario-3 differs from the one found in scenario-1.

\begin{table}[h!]
\centering
\begin{tabular}{|c|c|c|c|}
\hline
\textbf{Scenario}               & $J^*$ & Closest Material (confidence) & Area ($10^{-3} m^2$)\\ \hline
1  & 3.79 & AISI 4130 norm (89.76\%)  & [2,2,2,2,2,2] \\ \hline
2 & 2.95 & AISI 4130 norm  (89.76\%) & [3.2,2.6,2.6,3.2,1,1] \\ \hline
3 & 2.58  & AISI 1010 (90.69\%)   & {[}3.6,3.1,3.1,3.5,1.3,1.2{]}  \\ \hline
\end{tabular}
\caption{Cost constrained optimization of the truss in \cref{fig:midCantBC}}
\label{table:perf_threeCases_costCons_midCant}
\end{table}

For scenario-3, \cref{table:closestMatsCostConstraint} lists the three closest materials and their confidence values. 

\begin{table}[h!]
\centering
\begin{tabular}{|l|c|}
\hline
\multicolumn{1}{|c|}{Closest Material} & Confidence ($\gamma$)\\ \hline
1. AISI 1010 & 90.69\%    \\ \hline
2. 4032-T6 & 86.64\%    \\ \hline
3. AISI 1020 & 80.01\%    \\ \hline
\end{tabular}
\caption{Three closest materials under cost constraint.}
\label{table:closestMatsCostConstraint}
\end{table}

As an additional experiment, we replaced the cost constraint (\cref{eq:optimization_nn_costCons}) in the above problem with a mass constraint:

\begin{equation}
\label{eq:mat_optimization_nn_massCons}
  g_{m} \coloneqq
	\bigg(\frac{\rho}{M^*} \sum\limits_{k=1}^{N} A_k L_k\bigg) - 1 \leq 0 
\end{equation}
with $M^* = 40$kg; all other parameters being the same. The results are presented in \cref{table:massCons_midCant_threeCases}. Once again we observe that simultaneous optimization results in the best performance. Furthermore, when cost constrained was imposed, a steel alloy was chosen as the best material, whereas when a mass constrained is imposed, an aluminium alloy was chosen.

\begin{table}[h!]
\centering
\begin{tabular}{|c|c|c|c|}
\hline
\textbf{Scenario}               & $J^*$ & Closest Material (confidence) & Area ($10^{-3} m^2$) \\ \hline
1 & 10.5 & Al 380F die (91.63\%) & [2,2,2,2,2,2] \\ \hline
2 & 8.75 & Al 380F die (91.63\%) & [3,2.5,2.4,3.3,1,1] \\ \hline
3 & 8.56 & Al 2018 (94.16\%) & [3.1,2.4,2.5,3.1,1,1] \\ \hline
\end{tabular}
\caption{Mass constrained optimization of the truss in \cref{fig:midCantBC}}
\label{table:massCons_midCant_threeCases}
\end{table}

For scenario-3 under mass constraint, \cref{table:closestMatsMassConstraint} lists the three  closest materials and the confidence values. 

\begin{table}[h!]
\centering
\begin{tabular}{|l|c|}
\hline
\multicolumn{1}{|c|}{Closest Material} & Confidence ($\gamma$)\\ \hline
1. Al 2018 & 94.16\%    \\ \hline
2. Al 380F die & 90.64\%    \\ \hline
3. Al 2014-T4 & 90.13\%    \\ \hline
\end{tabular}
\caption{Three closest materials  under mass constraint.}
\label{table:closestMatsMassConstraint}
\end{table}

\subsection{Material Database Subsets}
\label{sec:expts_materialSubsets}

While we have used the entire material database in the previous experiment, we now consider subsets of the database. Subsets of materials can often be  justified based on experience. The VAE reconstruction error for the full database and three sample subsets is reported in \cref{table:vaeReconsErrorSubsets}. As one expect, the VAE error reduces for the subsets. 

\begin{table}[h!]
\centering
\begin{tabular}{|c|c|c|c|c|c|}
\hline
\textbf{Material class} & No. materials & \textbf{$\Delta E \%$} & \textbf{$\Delta C \%$} & \textbf{$\Delta \rho \%$} & \textbf{$\Delta Y \%$} \\ \hline
All Matls.  & 92 & 5.0 & 6.8 & 3.3 & 7.1 \\ \hline
Steel    & 14 & 0.58 & 2.6  & 0.07 & 1.4 \\ \hline
Aluminum & 52 & 0.09 & 0.22 & 0.04 & 2.9 \\ \hline
Plastic  & 12 & 2.3  & 0.36 & 0.5  & 0.7 \\ \hline
\end{tabular}
\caption{VAE reconstruction error trained with the full and various material subsets.}
\label{table:vaeReconsErrorSubsets}
\end{table}

We now carry out a few  experiments using these subsets, and report the results. First, for the truss problem in  \cref{fig:midCantBC}, we carry out a cost-constrained simultaneous optimization for the full database, as in the previous section. However,  we now report additional results in \cref{table:subsetCostConstraint}. When all materials are considered (row-2), the compliance prior to snapping to the closest material is reported as $J = 2.25$. The closest material is AISI 1010, resulting in a final compliance of $J^* = 2.58$. As one can expect, $J^* > J$.

Next, since  AISI 1010 is a steel alloy, we now consider only the  Steel subset, and carry out simultaneous optimization. The results are reported in  row-3. Observe that the raw compliance $J=2.6$ is surprisingly larger than $J = 2.25$ in the previous row. This  can be attributed to the larger VAE reconstruction error when all materials are considered, i.e., $J = 2.25$ is not as reliable as the estimate $J = 2.6$. Further, after the closest material is selected (AISI 1020), the final compliance is $J^* = 2.6$. This is slightly larger than  the value in the previous row. This discrepancy is analogous to rounding errors in integer optimization \cite{martins2021engineering}.

\begin{table}[h!]
\centering
\begin{tabular}{|c|c|c|c|c|c|c|c|}
\hline
\textbf{Material class} & $J$ & Optimal material & $E$ & $\rho$ & \textbf{$Y$} & $J^*$ & Area ($10^{-3}m^2$) \\ \hline
All Matls. & 2.25 & AISI 1010          & 2E11  & 7.9E3 & 3.3E8 & 2.58 & {[}3.6,3.1,3.1,3.5,1.3,1.2{]} \\ \hline
Steel     & 2.6  & AISI 1020          & 2E11  & 7.9E3 & 4.2E8 & 2.6  & {[}3.7,2.9,3.1,3.8,1.1,1.2{]} \\ \hline
\end{tabular}
\caption{Material refinement with cost constraint.}
\label{table:subsetCostConstraint}
\end{table}

Next, we repeated the above experiment for a mass constraint (instead of a cost constraint) and report the results in \cref{table:subsetMassConstraint}. When all materials are considered (row-2), the algorithm converges to the aluminum class, with an optimal material Al 2018, as before. Further, $J^* > J$ as expected. The results using the aluminum subset is reported in row-3. Now, the optimal material is AA356.0-F, and it differs from Al 2018. Now the performance $J^* = 8.38$ has significantly improved compared to  $J^* = 8.53$.

\begin{table}[h!]
\centering
\begin{tabular}{|c|c|c|c|c|c|c|c|}
\hline
\textbf{Material class} & $J$ & Optimal material  & $E$ & $\rho$ & $Y$ & $J^*$ & Area ($10^{-3}m^2$) \\ \hline
All Matls. & 8.21 & Al 2018     & 7.40E10 & 2.80E3 & 4.21E8 & 8.53 & {[}3.1,2.4,2.5,3.1,1,1{]}     \\ \hline
Aluminum & 8.23  & AA356.0-F        & 7.24E10 & 2.68E3 & 1.45E8 & 8.38 & {[}3.2,2.7,2.6,3.2,0.9,1{]}   \\  \hline

\end{tabular}
\caption{Material refinement with mass constraint.}
\label{table:subsetMassConstraint}
\end{table}

\subsection{Convergence}
\label{sec:expts_convergence}
\begin{figure}[h!]
    \centering
    \includegraphics[scale = 0.6]{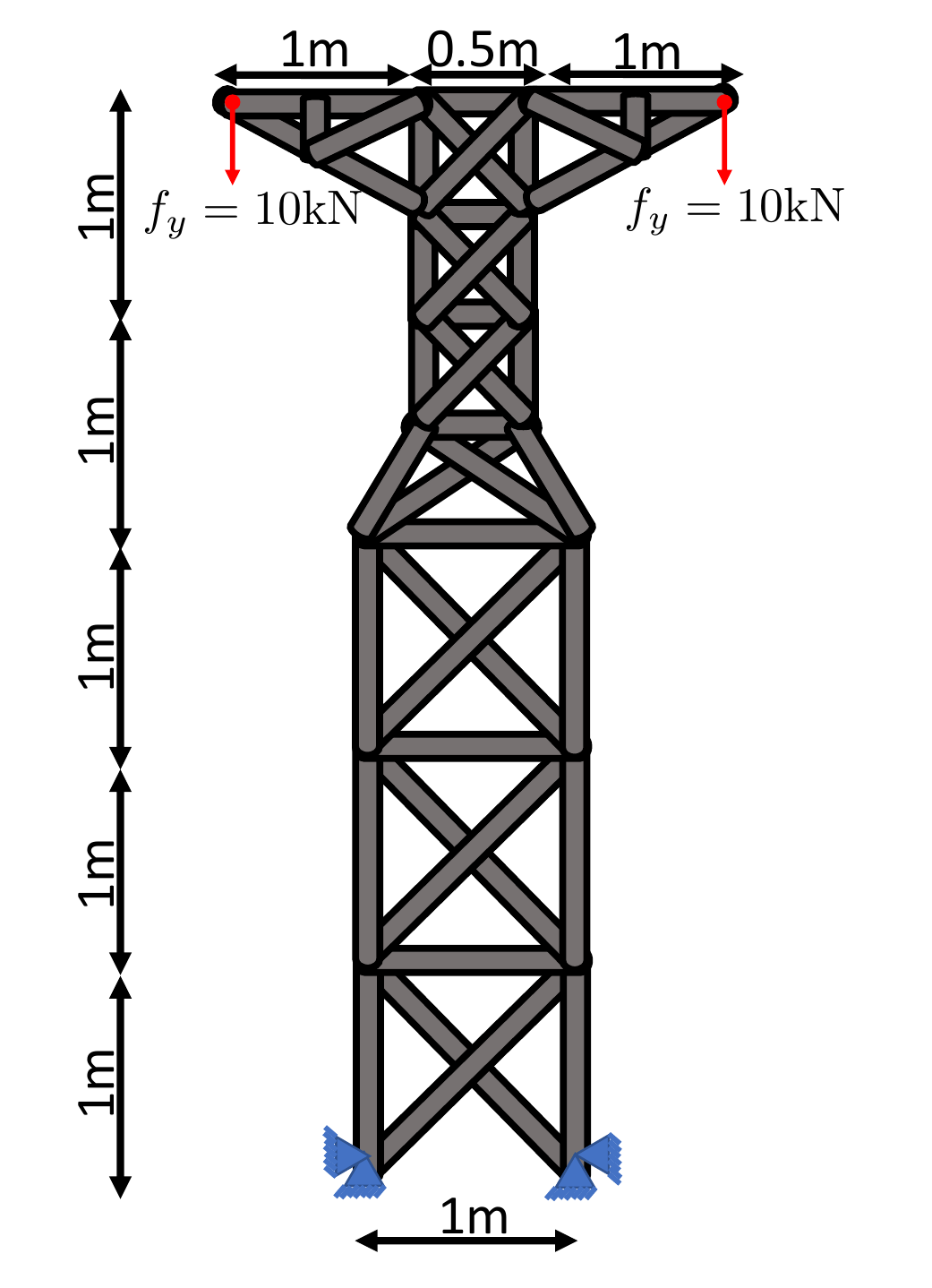}
    \caption{Loading of 47-bar antenna tower.}
    \label{fig:antennaBC}
\end{figure}

\begin{figure}[h!]
    \centering
    \includegraphics[scale = 0.65]{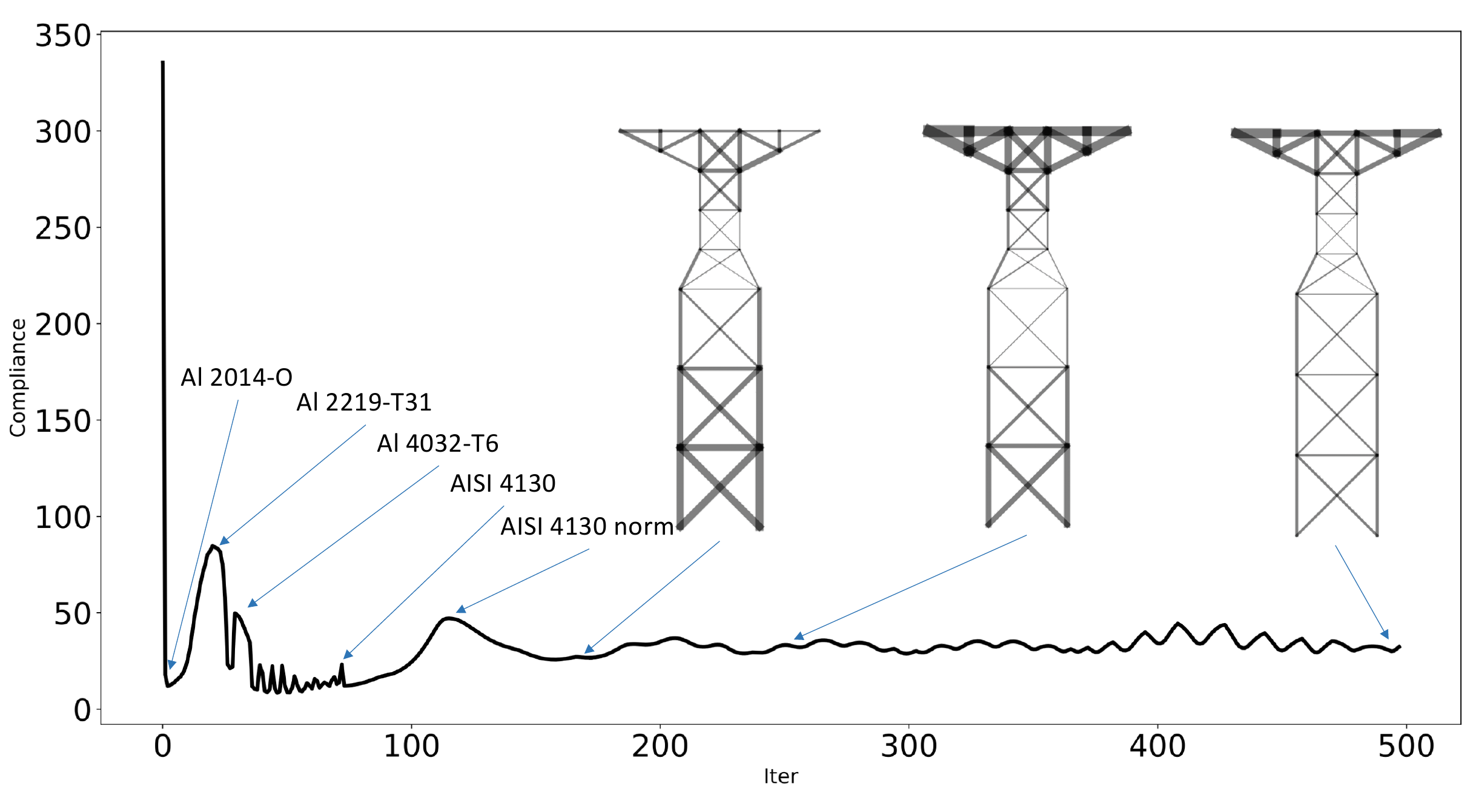}
    \caption{Convergence of compliance for the material and area for loading in \cref{fig:antennaBC}.}
    \label{fig:antenna_compliance}
\end{figure}

To study the convergence characteristics of simultaneous optimization, we consider the antenna tower illustrated in \cref{fig:antennaBC}. We impose a cost constraint of $\$50$, factor of safety  of $F_S = 4$, and area bounds $[10^{-9}, 10^{-2}] m^2$. Then the areas and material are simultaneously optimized, resulting in an optimal material of  `AISI 4130 norm' and $J^* = 32.5$.  The convergence in compliance is illustrated in \cref{fig:antenna_compliance}.  We observe that the material choice  converges early while the areas continue to be optimized; this was  typical. The optimization took $1.89$ seconds, with forward propagation and FEA, accounting for $16\%$ each, while back-propagation consumed $54\%$.

\section{Conclusion}
\label{sec:conclusion}

Engineers today have access to over $150,000$ materials \cite{ASHBY20111}, and this number is  growing as new materials are being discovered  \cite{ward2016general} \cite{ge2019accelerated}. This presents a significant challenge to design engineers. In this paper, using variational auto encoders (VAEs), we proposed a generic method to simultaneously select the optimal material, while optimizing the geometry. The proposed method was demonstrated using trusses.

There are several limitations to the proposed method. A heuristic confidence-metric was used, as a final step, to snap from the continuous latent space to the nearest material. While this is a well known method in integer optimization, and was found to be effective in out study, it might lead to discrepancies in performance, as was noted in the numerical experiments. On the other hand, since the proposed method suggests an optimal set of material properties, it  could drive material innovation in targeted applications. A very small material database with about 100 materials, with limited number of attributes, was considered here. A larger database and additional attributes needs to be explored.  The method could benefit from better tuning of the networks, optimal choice of the NN architecture, and increased latent space.

Despite these limitations, we believe that the method holds promise. For example, it could be extended to topology optimization with material selection, specifically using the NN framework proposed in \cite{ChandrasekharTOuNN2020}. Furthermore, inclusion of   thermo-elastic properties \cite{giraldo2020multi}, auxetic  properties \cite{takenaka2012negative},  coating properties \cite{clausen2015topologyOptCoating}, global warming indices \cite{ching2021truss} can also be considered. Inclusion of uncertainty in material properties is also of significant interest. Finally, the method can be extended to the selection of discrete components such as springs, bolts, etc., and  to the selection of discrete microstructures \cite{chan2021remixing}. 
\section*{Compliance with ethical standards}
The authors declare that they have no conflict of interest.

\section*{Reproduction of Results}
\label{sec:reprod}
The Python code pertinent to this paper is available at \href{https://github.com/UW-ERSL/MaTruss}{https://github.com/UW-ERSL/MaTruss}. The material data was sourced from SolidWorks.

\section*{Acknowledgements}
\label{sec:acknowledgements}
The authors would like to thank the support of National Science Foundation through grant CMMI 1561899.

\bibliographystyle{unsrt}
\bibliography{references}

\end{document}